\documentclass[aps,prl,preprint,superscriptaddress,showpacs]{revtex4}
\usepackage{epsfig}
\usepackage{latexsym}

\begin{document}     

\title{Large transverse Hall-like signal in topological Dirac semimetal $\rm Cd_3As_2$}

\author{Shih-Ting Guo}
\affiliation{Institute of Physics, Academia Sinica, Nankang, Taipei 11529, Taiwan}
\author{R. Sankar}
\affiliation{Center for Condensed Matter Sciences, National Taiwan University, Taipei 10617, Taiwan}
\affiliation{Institute of Physics, Academia Sinica, Nankang, Taipei 11529, Taiwan}
\author{Yung-Yu Chien}
\affiliation{Institute of Physics, Academia Sinica, Nankang, Taipei 11529, Taiwan}
\author{Tay-Rong Chang}
\affiliation{Department of Physics, National Tsing Hua University, Hsinchu 30013, Taiwan}
\author{Horng-Tay Jeng}
\affiliation{Department of Physics, National Tsing Hua University, Hsinchu 30013, Taiwan}
\affiliation{Institute of Physics, Academia Sinica, Nankang, Taipei 11529, Taiwan}
\author{Guang-Yu Guo}
\affiliation{Department of Physics, National Taiwan University, Taipei 10617, Taiwan}
\author{F. C. Chou}
\affiliation{Center for Condensed Matter Sciences, National Taiwan University, Taipei 10617, Taiwan}
\author{Wei-Li Lee}\email{wlee@phys.sinica.edu.tw}
\affiliation{Institute of Physics, Academia Sinica, Nankang, Taipei 11529, Taiwan}
\date{\today}      

\begin{abstract}

Cadmium arsenide ($\rm Cd_3As_2$) is known for its inverted band structure and ultra-high electron mobility. It has been theoretically 
predicted and also confirmed by ARPES experiments to exhibit a 3D Dirac semimetal phase containing degenerate Weyl nodes. From 
magneto-transport measurements in high quality single crystals of $\rm Cd_3As_2$, a small effective mass $m^* \approx$ 0.05 $m_e$ is 
determined from the Shubnikov-de Haas (SdH) oscillations. In certain field orientations, we find a splitting of the SdH oscillation 
frequency in the FFT spectrum suggesting a possible lifting of the double degeneracy in accord with the helical spin texture at outer and 
inner Fermi surfaces with opposite chirality predicted by our \textit{ab initio} calculations. Strikingly, a large antisymmetric 
magnetoresistance with respect to the applied magnetic fields is uncovered over a wide temperature range in needle crystal of $\rm 
Cd_3As_2$ with its long axis along [112] crystal direction. It reveals a possible contribution of intrinsic anomalous velocity term in the 
transport equation resulting from a unique 3D Rashba-like spin splitted bands that can be obtained from band calculations with the 
inclusion of Cd antisite defects.       
                   
\end{abstract}
\pacs{}
\maketitle
Topological materials have attracted great attention recently in condensed matter physics and material science. Non-trivial topology in a 
bulk band along with certain crystal symmetry can give rise to a novel material phase with unusual surface states, such as topological 
insulator \cite{TI1,TI2}, topological crystalline insulator \cite{TCI1,TCI2} and Weyl semimetal (WS) \cite{weyl1,weyl2,weyl3}. Recently, a 
3D Dirac semimetal (DS) phase has been theoretically predicted to exist in $\rm Na_3Bi$ \cite{dirac1}, $\rm BiO_2$ \cite{dirac2}, and $\rm 
Cd_3As_2$ \cite{dirac3}, where angle-resolved photoemission spectroscopy (ARPES) has provided direct evidences for such a 3D Dirac 
semimetal band \cite{ARPESdirac,neupane}. As opposed to a WS, a DS contains overlapping Weyl nodes with opposite chirality in momentum 
space and typically requires a special crystal symmetry to protect the nodes against the gap-opening. The breaking of either time reversal 
symmetry or inversion symmetry in a DS is, therefore, a route toward realizing a WS phase, where a special magneto-electric coupling 
effect \cite{magnetoelectric} due to the non-conserving chiral current between separated Weyl nodes can lead to novel transport phenomena 
\cite{son,chiralanomaly,ongcd3as2,burkov,li}. $\rm Cd_3As_2$ contains inverted bands with ultra-high electron mobility, which were 
reported decades ago \cite{cd3as21,cd3as22}, but only recently it revives as an example of 3D topological Dirac semimetals. It has been 
shown in experiments that a WS phase can be realized in $\rm Cd_3As_2$ by applying intense magnetic field along the [001] crystal 
direction \cite{jeon,cao}. However, when the magnetic field is tilted away from [001] direction, the four-fold rotational symmetry is 
broken giving rise to gapped Dirac nodes in $\rm Cd_3As_2$ \cite{xiang}. More recently, an exotic superconducting phase was discovered in 
point contact measurements on the surface of $\rm Cd_3As_2$, which was attributed to the possible tip-induced symmetry-lowering or density 
variation near the point contact region \cite{aggarwal,wang}. Those results all point to an important fact that the system's symmetry 
plays a crucial role dictating the details of Dirac band structures in $\rm Cd_3As_2$.      
  
In this study, we performed magnetotransport measurements on needle crystals of pure $\rm Cd_3As_2$ grown by chemical vapor transport 
\cite{sankar,feng}. Compared to flux growth single crystals \cite{ongcd3as2,zhao}, the CVT needle crystals have lower residual resistivity 
ratio suggesting higher defect level, which enables us to investigate the influence of defects to the Dirac band structures in $\rm 
Cd_3As_2$. By introducing the Cd antisite defects into the band calculations, both the inversion symmetry and rotational symmetry were 
broken giving rise to a unique 3D Rashba-like spin splitted bands in $\rm Cd_3As_2$, which may provide a qualitative explanation to the 
observed splitting in SdH frequency at various field angles and also a huge antisymmetric magnetoresistance (MR) with respect to magnetic 
fields in $\rm Cd_3As_2$. Nevertheless, a possible current-jetting effect \cite{jet1,jet2} due to disorder and inhomogeneous conductivity, 
which has been reported in several narrow band-gap semiconductors \cite{ag2se,insb}, will be discussed and compared to our transport data 
in $\rm Cd_3As_2$.

\subsection{Results and Discussions}
\begin{figure}[ht]
\centerline 
{\epsfig{figure=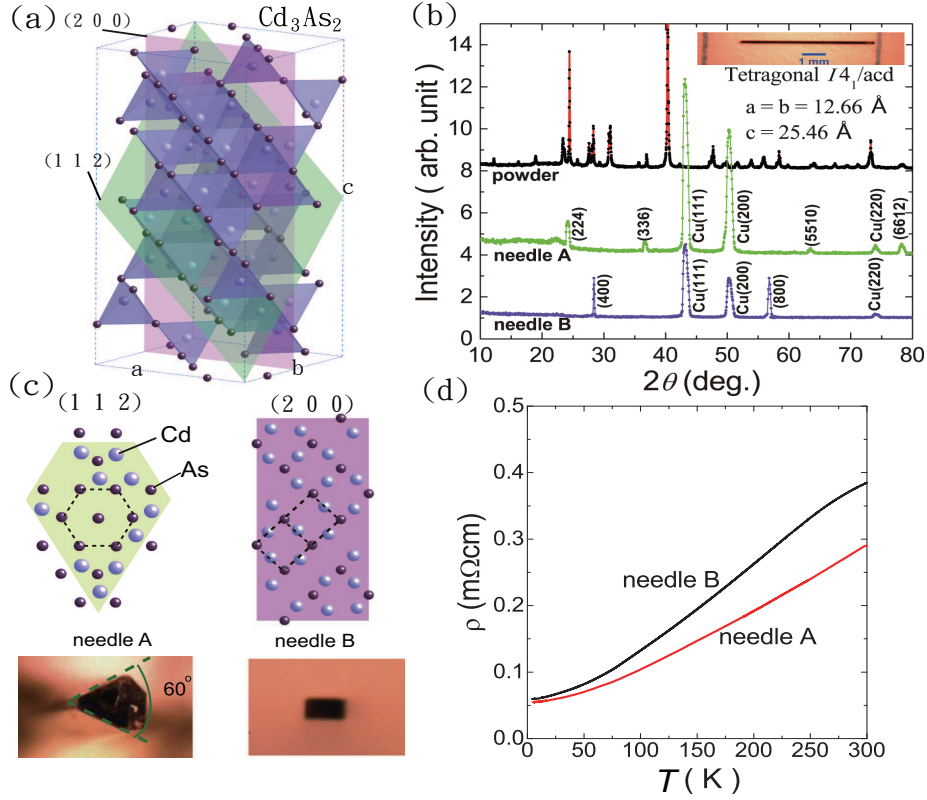,height=4.2in,width=4.8in,clip=0}}
\caption {\label{rt} (color online) (a) An illustration of $\rm Cd_3As_2$ crystal structure, where the (112) and (200) plane are shown. 
(b) Powder XRD pattern with preferred orientation fitted by I4$\rm_1$/acd space group symmetry. The single crystal XRD of needle A and B 
are also included to verify the corresponding needle long-axis directions to be along [112] and [200], respectively. The additional copper 
peaks come from the sample holder contribution. The upper inset shows the optical image of a 6 mm-long needle crystal of $\rm Cd_3As_2$. 
(c) Optical images of needle A and B, showing triangular and rectangle cross-sections, where the corresponding (112) and (200) planes are 
illustrated in the upper panel. (d) Temperature dependence of the resistivity for needle A and needle B, showing similar metallic 
behavior. } 
\end{figure}

The crystal structure of $\rm Cd_3As_2$ as shown in Fig. \ref{rt}(a) comprises a distorted antifluorite structure with ordered Cd 
vacancies, and it contains 80 atoms within a unit cell. For the phase stabilized at the lowest temperature, the Cd vacancies could be 
arranged orderly within a large tetragonal cell composed of corner-sharing $\rm CdAs_4$ tetrahedral units. The symmetry of $\rm Cd_3As_2$ 
generated at the lowest temperature has been controversially indexed with either noncentrosymmetric space group I4$\rm_1$cd 
\cite{steigmann} or I4$\rm_1$/acd with centrosymmetry \cite{Ali2013}. While the indexing based on powder X-ray diffraction is nearly 
equally satisfactory, and the initial Cd vacancy ordering is determinative on the final symmetry, the growth condition and crystal 
morphology could also play a role \cite{Ali2013}. In the same batch of crystals, two types of needles can be identified. Needle A has 
nearly triangular-shape cross-section, while needle B has rectangle-shape cross-section as demonstrated in Fig. \ref{rt}(c). Fig. 
\ref{rt}(b) shows the powder XRD pattern of the needle B crystal, which can be indexed with the space group I4$\rm_1$/acd with 
centrosymmetry. For needle A, the same space group of I4$\rm_1$/acd can be indexed in the X-ray diffraction. In order to identify the 
crystalline direction along the long axes of the needle crystals, a 4-circle diffractometer was used, and the corresponding single-crystal 
XRD of needle A and B are shown in Fig. \ref{rt}(b), where the long axis directions of needle A and B were confirmed to be along [112] and 
[200], respectively. In addition, the full-width half maximum of the XRD peaks is merely about 0.2 - 0.4 degrees indicating good single 
crystalline quality with a small mosaic spread in our needle-like crystals. Crystallographically, it is reasonable to have stable phase of 
large (112) plane for $\rm Cd_3As_2$, because (112) plane corresponds to a plane of pseudo-hexagonal close packing for a tetragonal unit 
cell with c $\approx$ 2a. The resistivity data shown in Fig. \ref{rt}(d) indicate a metallic behavior with residual resistivity ratios 
(RRR$\equiv\rho\rm_{300K}/\rho\rm_{5K}$) of 4.4 and 5.6 for needle A and needle B, respectively. The corresponding electron density and 
the estimated Drude mobility are listed in Table \ref{para}. For needle B, the carrier density equals 9.2$\rm\times 10^{17} cm^{-3}$, and 
the corresponding Hall mobility $\mu_D$ is as high as 113,567 cm$^2$V$^{-1}$s$^{-1}$ at $T$ = 5 K.
  
\begin{figure}[ht]
\centerline 
{\epsfig{figure=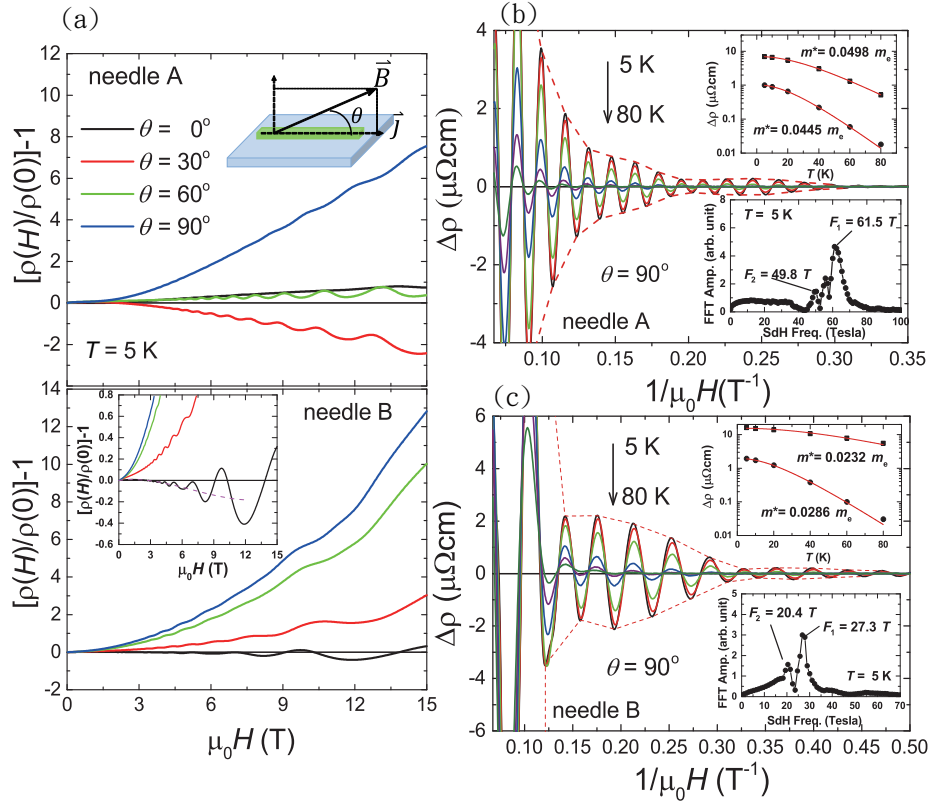,height=4.2in,width=4.8in,clip=0}}
\caption {\label{sdh} (color online) (a) Symmetrized MR in needle A (upper) and B (lower) at four different $\theta$ angles. The inset 
figure of lower panel is an enlarged view, showing a small negative MR at $\theta$ = 0$\rm^o$. (b) and (c) plot the pure oscillatory 
component in MR versus 1/$\mu_0H$ for needle A and B, respectively, at $\theta$ = 90$\rm^o$ and various temperatures ranging from 5 to 80 
K. The upper and lower insets are the corresponding effective masses fitting using Lifshitz-Kosevich formula and FFT spectra. Both needle 
A and B exhibit beating patterns in SdH oscillations with multiple closely-spaced peaks in the FFT spectra at $\theta$ = 90$\rm^o$. } 
\end{figure}

Figure \ref{sdh}(a) shows the symmetrized MR (MR = [MR($H$) + MR(-$H$)]/2) at four different $\theta$ values ranging from zero to 90 
degrees, where $\theta$ is defined as the angle between the current and the applied magnetic field (upper inset cartoon). At $T$ = 5 K, 
the magnitude of MR in needle B progressively decreases from a large positive value of MR $\equiv [\rho(H)/\rho(0)]-1$ = 12.8 at $\theta$ 
= 90$\rm^o$ to a small negative MR at $\theta$ = 0$\rm^o$ angle (shown in the lower inset of Fig. \ref{sdh}(a)). On the contrary, the MR 
in needle A shows non-monotonic variation with $\theta$ values as demonstrated in the upper panel of Fig. \ref{sdh}(a). The unusual high 
$\mu_H$ enables the determination of band parameters via SdH oscillations in transport measurements. A large amplitude of SdH oscillation 
was found in MR at low $T$, which remains observable up to $T \geq$ 100 K. Figure \ref{sdh}(b) shows the pure oscillatory component of the 
resistivity $\Delta\rho$ versus 1/$\mu_0H$ for needle A at 6 different temperatures ranging from 5 K to 80 K. The magnetic field is normal 
to the [112] direction. The damping of the SdH oscillation by temperature and magnetic field can be described by the following equation 
based on Lifshitz-Kosevich formula.
\begin{equation}
\Delta\rho(T,B) = {\rm exp}[-X(T_D,B)]\frac{X(T,B)}{{\rm sinh}(X(T,B))}\Delta\rho^\prime,
\label{lkformula}
\end{equation}  
where $X (T,B)\equiv 2\pi^2k_BTm^*/\hbar eB$, $m^*$ is the effective cyclotron mass and $T_D$ is the Dingle temperature. 
$\Delta\rho^\prime$ refers to the undamped oscillatory component. By fitting the peak values of $\Delta\rho$-$T$ and 
log($\rm\Delta\rho\cdot sinhX/X$)-1/$\mu_0H$ according to Eq. \ref{lkformula}, we determined $m^*$ = 0.0498 $m_e$ for needle A as 
demonstrated in the upper inset of Fig. \ref{sdh}(b). Two resolvable SdH frequencies of 49.8 and 61.5 Tesla were identified in the Fast 
Fourier transform (FFT) spectrum shown in the lower inset of Fig. \ref{sdh}(b). On the other hand, the pure oscillatory component of 
resistivity for needle B is shown in Fig. \ref{sdh}(c), where two SdH frequencies of 20.4 and 27.3 Tesla and $m^*$ = 0.0232 $m_e$ were 
determined. The calculated band for I4$\rm_1$/acd structure with Cd antisites in (112) plane at similar $n_e$ gives two close SdH 
frequencies $F_1$ and $F_2$, which is in relatively good agreement with the experimental data. A summary of the band parameters for needle 
A and needle B from SdH experiments and calculation is given in Table \ref{para}. 

\begin{table}
\caption{Major SdH frequency $\it{F_1}$(T), secondary SdH frequency $\it{F_2}$(T) effective mass $m^*(m_e)$ and Drude mobility 
$\mu_D$(cm$^2$V$^{-1}$s$^{-1}$) obtained from SdH oscillation measurements and calculated band structure.} 
\centering 
\begin{tabular}{|l|l|l|l|l|l|} 
\toprule
&$n_e\times$10$^{18}$&\it{F$_1$}&\it{F$_2$}&$m^*_1$&$\mu_D$ at 5 K \\ 
& \scriptsize{(cm$^{-3}$)}&\footnotesize{(T)}& \footnotesize{(T)}& \footnotesize{($m_e$)}& \footnotesize{(cm$^2$V$^{-1}$s$^{-1}$)}\\
\colrule 
\footnotesize{needle A} & 2.4 &61.5&49.8 & 0.0498& 45,577 \\ 
\footnotesize{needle B} & 0.92 &27.3&20.4 &0.0232& 113,567\\
\colrule 
Calculation &2.4&52 &35&-&-\\
&0.92&31 &18&-&-\\
\botrule 
\end{tabular}
\label{para}
\end{table}  

\begin{figure*}[ht]
\centerline 
{\epsfig{figure=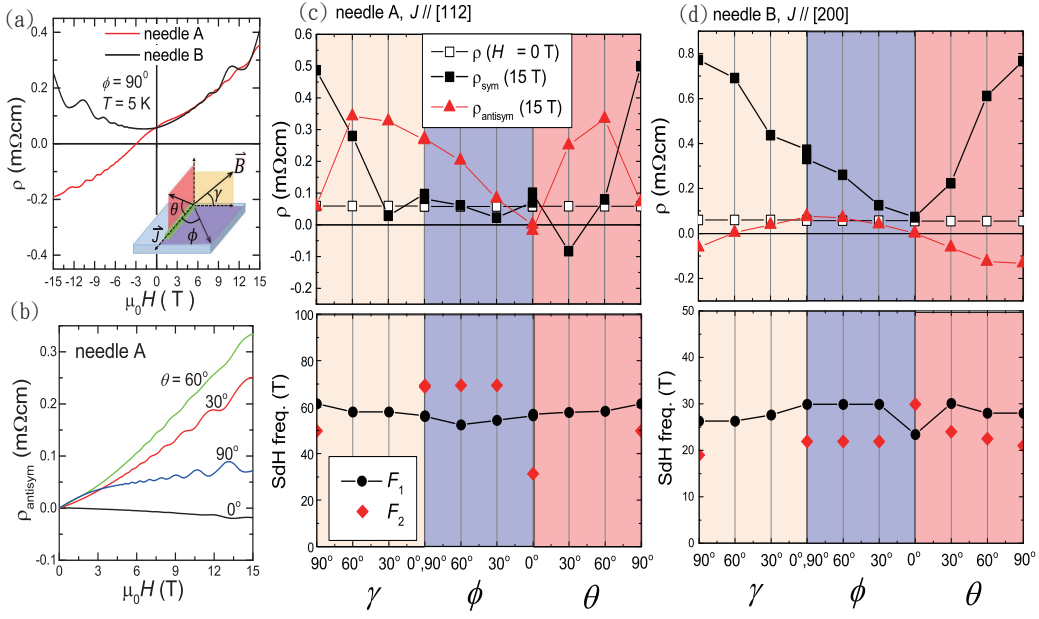,height=3.2in,width=5.4in,clip=0}}
\caption {\label{angle} (color online) (a) Resistivity $\rho$ versus field $\mu_0H$ shows significant antisymmetric component in needle A 
and B. The lower inset cartoon illustrates the geometric definition for angles of $\gamma$, $\phi$, and $\theta$. (b) Extracted 
antisymmetric components ($\rho_{antisym}$) in MR of needle A at four different $\theta$ angles. At $\mu_0H$ = 15 Tesla, the angular 
dependence of extracted symmetric ($\rho_{sym}$) and antisymmetric ($\rho_{antisym}$) components of needle A and B are shown in (c) and 
(d), respectively. The lower panel shows the corresponding SdH frequencies that can be identified from the FFT spectra. The major SdH 
frequencies $F_1$ (black circles) at about 61 Tesla and 27 Tesla for needle A and B, respectively, both show weak angular dependence. 
Secondary SdH frequencies $F_2$ appear at some angles shown as red diamonds.} 
\end{figure*}

Figure \ref{angle}(a) shows the resistivity versus magnetic field applied along $\phi$ = 90$\rm^o$, where the definition of corresponding 
angles of $\gamma$, $\phi$, and $\theta$ are illustrated in the lower inset cartoon. Both needle A and needle B exhibit noticeable 
antisymmetric component in the MR regardless of the large aspect ratio ($\equiv$ length/width) of $\cong$ 3.9 (15.7) in needle A (B). By 
using the formula of $\rho_{sym} \equiv [\rho(H)+\rho(-H)]/2$ and $\rho_{antisym} \equiv [\rho(H)-\rho(-H)]/2$, symmetric ($\rho_{sym}$) 
and antisymmetric ($\rho_{antisym}$) components of the MR can be extracted. The resulting field dependence of $\rho_{antisym}$ in needle A 
at four different $\theta$ angles is shown in Fig.\ref{angle}(b), where the corresponding $\rho_{sym}$ has been shown in Fig. 
\ref{sdh}(a). The upper panels of Fig. \ref{angle}(c) and \ref{angle}(d) are the angular dependence of the extracted $\rho_{sym}$ and 
$\rho_{antisym}$ at $\mu_0H$ = 15 T for needle A and needle B, respectively. We remark that, for needle B with current along the [200] 
crystal direction, $\rho_{antisym}$ at some angles can be even larger than $\rho_{sym}$, and the magnitude of $\rho_{antisym}$ appears to 
be at minimum when $\phi$ = 0$\rm^o$ or $\theta$ = 0$\rm^o$. The SdH frequencies determined from the FFT spectra for needle A and needle B 
are shown in the lower panels of Fig. \ref{angle}(c) and \ref{angle}(d), respectively, where apparent multiple SdH frequencies can be 
identified. The major SdH frequencies $F_1$(black circles) being the location of largest peak in the FFT spectrum were determined to be 
about 61 Tesla and 27 Tesla in needle A and needle B, respectively, exhibiting weak dependency on $\gamma$, $\phi$, and $\theta$ values. 
We also note that secondary SdH frequencies (red diamonds) can be clearly observed at some angles in both needle A and needle B.         

\begin{figure}[ht]
\centerline 
{\epsfig{figure=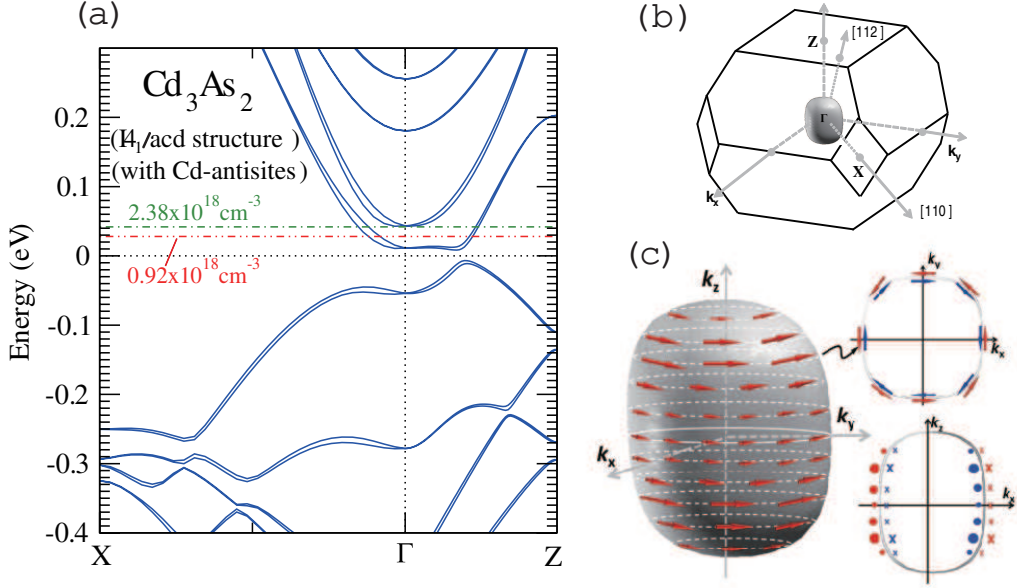,height=3.2in,width=5.4in,clip=0}}
\caption {\label{band} (color online) (a) A calculated band structure of Cd$_3$As$_2$ with I4$\rm_1$/acd space group symmetry and Cd 
antisite defects. The two dotted lines correspond to the Fermi level locations for needle A and B, respectively, determined from the 
experimental values of Fermi surface area. (b) The calculated 3D Fermi surface with an energy of $\approx$ 0.04 eV above the gapped Dirac 
node close to the Fermi level of our samples suggests a slightly distorted Fermi sphere in $\rm Cd_3As_2$. [112] and [200] crystal 
directions are indicated corresponding to the long-axis directions for needle A and B, respectively. (c) The spin texture over an outer 
Fermi surface. Red and blue arrows indicate spin texture on the outer and inner Fermi surfaces, respectively, where the size of the arrows 
indicates the spin polarization. The lower inset shows the spin texture profile on $k_x$-$k_z$ plane, while the upper inset is a top view 
of the Fermi surfaces at finite $k_z$ indicated by the black arrows.  } 
\end{figure}

Figure \ref{band}(a) shows the calculated band structure based on I4$\rm_1$/acd space group symmetry with Cd antisite defects. The 
dotted-dash lines represent two Fermi level locations set by experimental SdH frequencies within the rigid band approximation. For needle 
B with a density of $n_e$ = 0.92 $\times$ 10$^{18}$ cm$^{-3}$, the Fermi level is about 30 meV above the Dirac node. As described 
previously, the Cd vacancies are ordered in the manner to keep the inversion symmetry in the defect-free I4$\rm_1$/acd lattice. The 
inversion symmetry then leads to the spin degenerate Dirac bands, which can not explain the observed splitted SdH frequencies. Inspired by 
our calculated spin splitted Dirac bands of the noncentrosymmetric I4$\rm_1$/cd lattice (see supplementary information), we thus introduce 
$\approx$ 1 \% Cd antisite defect in the centrosymmetric I4$\rm_1$/acd lattice by moving one Cd ions to one of the Cd vacancies so as to 
break the inversion symmetry and split the centrosymmetry protected spin degenerate Dirac bands. As shown in the figure, a lifting of spin 
degeneracy in the calculated band in the $k_x$-$k_y$ plane is successfully obtained from breaking of the inversion center of I4$\rm_1$/acd 
structure by the Cd antisite defects and thus gives rise to the two calculated SdH frequencies listed in Table \ref{para}. This is in 
accord with the observed beating patterns and multiple closely-spaced SdH frequencies shown in Fig. \ref{sdh}(b) and (c) (see also 
supplementary information). However, we note that a small gap about 10 meV is opened at the Dirac node due to the rotational symmetry 
breaking by the Cd antisites. In addition, the band width was reduced causing a flatter Dirac bands, which results in a somewhat lower 
Fermi level location in our band calculations compared to previous reports based on defect-free band calculations \cite{zhao,neupane}. In 
the lower panels of Fig. \ref{angle}(c) and (d), the variation  of the SdH Frequency with $\gamma$, $\theta$, and $\phi$ is less than 15 
\%. The Fermi surface can, therefore, be regarded as a slightly distorted sphere as illustrated in Fig. \ref{band}(b) obtained from a 
calculated band with a Fermi level at $\approx$ 0.04 eV above the gapped Dirac node for the I4$\rm_1$/acd structure with $\approx$ 1 \% Cd 
antisite defects. The shape of the Fermi surfaces and the spin chirality are similar to those of the defect-free noncentrosymmetric 
I4$\rm_1$cd lattice at 0.1 eV with smaller band splittings. This remains well above the Lifshitz transition occurring at around 0.01 eV. 
Hence, the two Fermi pockets corresponding to two gapped Dirac nodes along the $\Gamma$-Z line merged together into a bigger Fermi surface 
centered at $\Gamma$. The [112] and [200] directions corresponding to the long axes of needle A and needle B, respectively, are also 
indicated in Fig. \ref{band}(b). We further performed calculation on spin texture of the 3D Fermi surface shown in Fig. \ref{band}(c), 
where the electrons over the Fermi surfaces possess chirality with a spin-momentum lock in-plane spin texture. Furthermore, due to the 
broken centrosymmetry by the Cd antisites, the spin texture on the inner and outer Fermi surfaces revolve in the opposite direction as 
shown in the upper inset of Fig. \ref{band}(c) similar to a Rashba-like band-splitting. The difference in the area of the two Fermi 
surfaces over the ab-plane is about 29\%, which is compatible with the experimental value ($\approx$ 20\%) from splitted SdH frequencies 
shown in Fig. \ref{sdh}(b) and (c). We remark that such an unusual 3D Rashba-like spin splitted band is a direct consequence of the large 
spin-orbit coupling and also Cd-antisites induced inversion symmetry breaking in $\rm Cd_3As_2$ while the time reversal symmetry remains 
valid.    

The raw data of the resistivity shows an antisymmetric behavior with respect to the in-plane magnetic field inferring a large transverse 
Hall-like signal picked up by the voltage leads due to the leads misalignment. The estimation of normal Hall signal due to leads 
misalignment is difficult due to unknown distance $W^\prime$ between leads normal to the current direction. A rough estimation of normal 
Hall contribution can be given using $\rho_{xy,norm} = \rho_{xy}\times(W^\prime/W)/(AR)$, where W is the needle width and AR = L/W is the 
aspect ratio of the needle. If we take $W^\prime$/W $\approx$ 10 $\%$ and AR = 3.9 (15.7) for needle A(B), $\rho_{xy, norm}$ (15 T) = 0.1 
m$\Omega$cm (0.06 m$\Omega$cm), which is at least 4-fold smaller in magnitude compared to the observed $\rho_{antisym}$ shown in Fig. 
\ref{angle}(c) and (d). In addition, $\rho_{xy, norm}$ should grow monotonically with increasing magnetic field strength normal to the 
current direction (i.e., the $\theta$ and $\phi$ angles go from zero to 90 degrees), which contradicts with the non-monotonic variation of 
$\rho_{antisym}$ with $\theta$ values observed in needle A shown in Fig. \ref{angle}(c).

Now we turn to discuss the possible current-jetting effect. It was shown that the inhomogeneous conductivity can cause a distortion in the 
current path giving rise to a perpendicular current component $\vec{J}_\perp$ that flows normal to the major current direction $\vec{J}$ 
\cite{jet2,ag2se}. In $\rm Cd_3As_2$, high resistivity anisotropy and relatively short quantum scattering lifetime of $\sim$ 10$^{-14}$ s 
as determined from SdH oscillations were reported \cite{ongcd3as2,ando} making the current jetting effect a likely source of unusual 
magnetotransport behavior. In general, $\vec{J}_\perp$ contributes a normal Hall signal to MR resulting in a $B$-linear and non-saturating 
MR in transverse field with $\vec{B} \perp \vec{J}$. For longitudinal field geometry of $\vec{B} \parallel \vec{J}$, further bending in 
$\vec{J}_\perp$ by $\vec{B}$ results in a decrease in potential drop between voltage electrodes and hence a negative MR. Compared to the 
MR data of $\rm Cd_3As_2$ shown in Fig. \ref{sdh}(a), a nearly $B$-linear MR at $\theta$ = 90$\rm^o$ ($\vec{B} \perp \vec{J}$) only 
appeared at fields higher than about 10 Tesla. In addition, a positive MR at $\theta$ = 0$\rm^o$ ( $\vec{B} \parallel \vec{J}$) with 
vanishing $\rho_{antisym}$ up to 15 Tesla (Fig. \ref{angle}(b)) was observed in needle A instead. Those phenomena cannot be fully 
understood via the current-jetting mechanism alone. We also note the large negative MR found in needle A at $\theta$ = 30$\rm^o$, where 
its origin is unclear and requires further investigation. On the other hand, $\rho_{antisym}$ remains finite at low-field diffusive limit 
and grows larger in higher fields without saturation as shown in Fig. \ref{angle}(b). In particular, the $\rho_{antisym}$-$\mu_0H$ curve 
at $\theta$ = 90$\rm^o$ exhibits a slower increase of $\rho_{antisym}$ at higher fields, which is different from others being nearly 
linear with field. Those findings strongly suggests that there is additional contribution to the antisymmetric MR other than the normal 
Hall contribution and the high field effect in quantum limit. We also find that such an antisymmetric behavior in MR persists up to room 
temperature and shows relatively weak $T$ dependence (see supplementary information), which supports for a more intrinsic origin rather 
than due to impurity effect at low $T$.

Another likely mechanism is the anomalous velocity term associated with the non-zero Berry curvature of the bulk band \cite{kla,klb}. In a 
general formalism, the electron velocity can be expressed as $\hbar\vec{\upsilon} = \nabla_{k}\epsilon(k) + e 
\vec{E}\times\vec{\Omega}(\vec{B},\vec{k})$, where $\vec{\Omega}(\vec{B},\vec{k})$ is the Berry curvature. In a diffusive transport 
\cite{son}, it can be shown to give a total current of 
\begin{equation}
\vec{J} = \int{\frac{d^3\vec{k}}{(2\pi)^3}[\vec{\upsilon} + e\vec{E}\times\vec{\Omega}(\vec{B},\vec{k}) + 
\frac{e}{c}(\vec{\Omega}(\vec{B},\vec{k})\cdot\vec{\upsilon})\vec{B}]n_{\vec{k}}},
\label{totalI}
\end{equation}   
where $n_{\vec{k}} = f^0_{\vec{k}} + g_{\vec{k}}(\vec{B}, \vec{E})$ is the electron distribution function. $f^0_{\vec{k}}$ and 
$g_{\vec{k}}(\vec{B}, \vec{E})$ are the equilibrium and non-equilibrium distribution function, respectively. In zero field, time reversal 
symmetry ensures that the integration of anomalous velocity term ($e \vec{E}\times\vec{\Omega}(\vec{B},\vec{k})$) with $f^0_{\vec{k}}$ 
over the whole Fermi surface gives zero contribution to the total current. However, in the presence of magnetic field and electric bias, 
the induced $g_{\vec{k}}(\vec{B}, \vec{E})$ can result in non-vanishing anomalous velocity contribution to the total current as shown in 
Eq. \ref{totalI}. It is known that the electric current and spin polarization can couple with each other in a 2D electron gas system with 
Rashba spin splitting \cite{2drashba, 2drashba2}. For example, current-induced spin polarization has been demonstrated providing a strong 
support for the observable effects originating from non-equilibrium $g_{\vec{k}}(\vec{B}, \vec{E})$ term. In the case of a 3D Rashba-like 
spin splitted band shown in Fig. \ref{band}(c), $g_{\vec{k}}(\vec{B}, \vec{E})$ value can have nontrivial dependence on the orientation of 
$\vec{B}$ and $\vec{E}$, which results in anomalous angular dependence in the transport phenomena. 

\begin{figure}[ht]
\centerline 
{\epsfig{figure=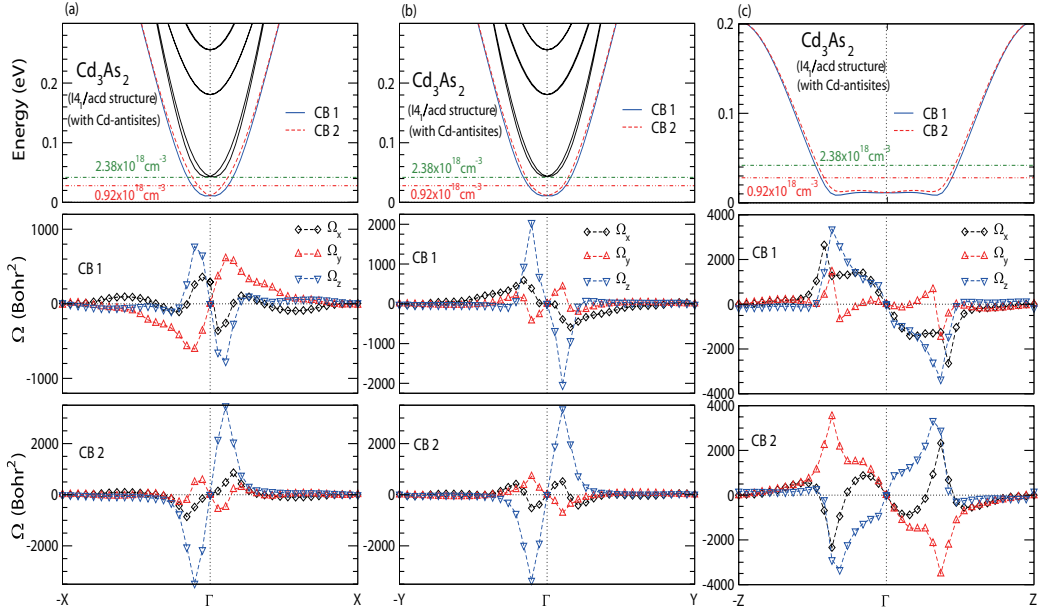,height=3.2in,width=5.4in,clip=0}}
\caption {\label{berry_curvature} (color online) Berry curvature calculations along $\Gamma-X$ (a), $\Gamma-Y$ (b), and $\Gamma-Z$ (c) 
using the band structure show in Fig. \ref{band}(a). The upper panels show the band structures. The corresponding Berry curvature 
$\vec{\Omega}$ for CB 1 and CB 2 bands are shown in the middle and lower panels, respectively. Non-zero Berry curvatures were found along 
all three principle axes in accord with a 3D nature of the Rashba-like spin-splitted band in $\rm Cd_3As_2$ with Cd antisites.  } 
\end{figure}

In order to look into the anomalous velocity contribution, Berry curvature $\vec{\Omega}$ based on the band structure shown in Fig. 
\ref{band}(a) was calculated. Figure \ref{berry_curvature}(a), (b), and (c) shows the results along $\Gamma-X$, $\Gamma-Y$, and 
$\Gamma-Z$, respectively. We note that $\vec{\Omega}$ for the two conduction bands of CB1 (middle panel) and CB2 (lower panel) have 
non-zero magnitudes along all three principle axes indicating a 3D nature of the Rashba-like spin-splitted band. $\vec{\Omega}$ is 
antisymmetric with respect to the $\Gamma$ point as expected for a system with time reversal symmetry. We remark that the calculated 
$\vec{\Omega}$ gives the largest non-zero value along $\Gamma-Z$ direction, which is nearly 4-fold bigger compared to that along 
$\Gamma-X$. Such a difference can provide qualitative explanation on the observation of a larger $\rho_{antisym}$ in needle A sample with 
current applied along [112], where a non-zero electric bias along $\Gamma-Z$ is present. We note that similar splitting in SdH frequency 
was also reported in flux-growth $\rm Cd_3As_2$ crystals \cite{zhao}. It was attributed to the two nested ellipsoidal Fermi surface along 
the $\Gamma-Z$, showing single SdH frequency when the magnetic field is along [112] direction, which is not compatible with our 
observation in needle A that shows two distinct SdH frequencies at $\theta$ = 0$\rm^o$. It is also not clear how the spin-degenerate and 
nested ellipsoidal Fermi surface can lead to a large transverse current response in the charge transport.       

In conclusion, $\rm Cd_3As_2$ exhibits a 3D topological Dirac semimetal phase with ultra-high electron mobility. According to our band 
calculations, Cd antisite defects can be an effective symmetry-breaking mechanism giving rise to a unique 3D Rashba-like spin splitting in 
the $k_x$-$k_y$ plane. This is also supported by our angular SdH measurements, showing splitted Fermi surface. By comparing the transport 
data in needle A and B with long axes along [112] and [200], respectively, we uncover significant transverse Hall-like signals in MR, 
which can not be simply attributed to the normal Hall-effect-related mechanisms. Particularly, we found that such a transverse Hall-like 
signal is much more pronounced in needle A with current bias along [112] direction. This can be qualitatively understood from our 
calculated Berry curvature, showing a 4-fold enhancement in magnitude along $\Gamma-Z$, and the corresponding non-equilibrium electron 
distribution due to the external magnetic field and electric bias can lead to an unusual large anomalous velocity contribution to the 
electron transport.   
         

\subsection{Acknowledgements}
The authors acknowledge the funding support from the nanoprogram at Academia Sinica and Ministry of Science and Technology (MOST) in 
Taiwan. W.L.L. acknowledges the funding support from the Academia Sinica 2012 career development award in Taiwan. G.Y.G. acknowledges the 
support from the Academia Sinica thematic research program. H.T.J. and G.Y.G. also thank NCHC, CINC-NTU and NCTS, Taiwan for supports. 
F.C.C. acknowledges support from TCECM under MOST-Taiwan. 
\subsection{Authors Contributions}
W.L.L., H.T.J., G.Y.G. and F.C.C. designed the experiment. S.R. and F.C.C. are responsible for single-crystal growth and crystal 
structural characterizations. T.R.C., H.T.J. and G.Y.G. performed the theoretical band calculations. S.T.G., Y.Y.C. and W.L.L. carried out 
the low temperature transport measurements and data analysis. W.L.L. wrote the manuscript.  
\subsection{Additional Information}
\textbf{Supplementary information} accompanies this paper at http://www.nature.com/srep
\textbf{Competing financial interests:} The authors declare no competing financial interests.

\clearpage

\end{document}